\documentclass[runningheads,a4paper]{llncs}
\usepackage{amsmath}
\usepackage{algorithm}
\usepackage[noend]{algpseudocode}
\usepackage{url}
\usepackage{psfrag}
\usepackage{epsfig}
\usepackage{verbatim}
\usepackage{amssymb}

\begin{document}

\title{Balancing the Communication Load of Asynchronously Parallelized Machine 
Learning Algorithms}

\titlerunning{Asynchronously Parallelized Machine Learning Algorithms}
\toctitle{Asynchronously Parallelized Machine Learning Algorithms}

\author{Janis Keuper and Franz-Josef Pfreundt}
\authorrunning{J. Keuper and F.-J. Pfreundt}
\tocauthor{J. Keuper and F.-J. Pfreundt}
\institute{Fraunhofer ITWM\\
       Competence Center High Performance Computing\\
       Kaiserslautern, Germany
\email{\{janis.keuper | franz-josef.pfreundt\}@itwm.fhg.de}
}

\maketitle
\begin{abstract}
Stochastic Gradient Descent (SGD) is the standard numerical 
method used to solve the core optimization problem for the vast majority of 
machine learning (ML) algorithms. In the context of large scale learning,
as utilized by many Big Data applications, efficient parallelization
of SGD is in the focus of active research.\\      
Recently,  we were able to show that  
the asynchronous communication paradigm can be applied to achieve a fast and 
scalable parallelization of SGD.  
Asynchronous Stochastic
Gradient Descent (ASGD) outperforms other, mostly MapReduce based,
 parallel algorithms solving large scale machine learning problems.\\
In this paper, we investigate the impact of asynchronous communication 
frequency and message size on the performance of ASGD applied to large scale ML on 
HTC cluster and cloud environments. We introduce a novel algorithm for the automatic 
balancing of the asynchronous communication load, which allows to adapt ASGD
to changing network bandwidths and latencies.     
\end{abstract}

\section{Introduction}
The enduring success of Big Data applications, which typically includes 
the mining, analysis and inference of very large datasets, is leading to a change 
in paradigm for machine learning research objectives \cite{bottou2008tradeoffs}. 
With plenty data at hand, the traditional challenge of inferring generalizing 
models from small sets of available training samples moves out of focus. Instead,
the availability of resources like CPU time, memory size or network bandwidth 
has become the dominating limiting factor for large scale machine learning
algorithms.\\
In this context, algorithms which guarantee useful results even in the case
of an early termination are of special interest. With limited (CPU) time,
fast and stable convergence is of high practical value, especially when the 
computation can be stopped at any time and continued some time later when more
resources are available.\\   
Parallelization of machine learning (ML) methods has been a 
rising topic for some time (refer to \cite{datamining} for a comprehensive
overview). Most current approaches rely on the MapReduce pattern. 
It has been shown \cite{chu2007map}, that most of the existing ML techniques
could easily be transformed to fit the MapReduce scheme. However,
it is also known \cite{recht2011hogwild}, that MapReduce's easy 
parallelization comes at the cost of potentially poor scalability.
The main reason for this undesired behavior 
resides deep down within the numerical properties most machine learning 
algorithms have in common: an optimization problem. In this context, 
MapReduce works very well for the implementation of so called batch-solver
approaches, which were also used in the MapReduce framework of 
\cite{chu2007map}. However, 
batch-solvers have to run over the entire dataset to compute a single 
iteration step. Hence, their scalability with respect to the data size is
obviously poor.\\
Therefore, even most small scale ML implementations 
avoid the known drawbacks of batch-solvers by usage of alternative online
optimization methods. Most notably, Stochastic Gradient Descent (SGD) methods 
have long  proven to provide good results for ML optimization problems \cite{recht2011hogwild}.\\
However, due to its inherent sequential nature, SGD is hard to parallelize and even harder 
to scale \cite{recht2011hogwild}.
 Especially, when communication latencies are causing dependency locks,
which is typical for parallelization tasks on distributed memory systems \cite{SGDsmola}.\\ 

\begin{figure}[!htbp]
\centering
\includegraphics[width=0.85\linewidth]{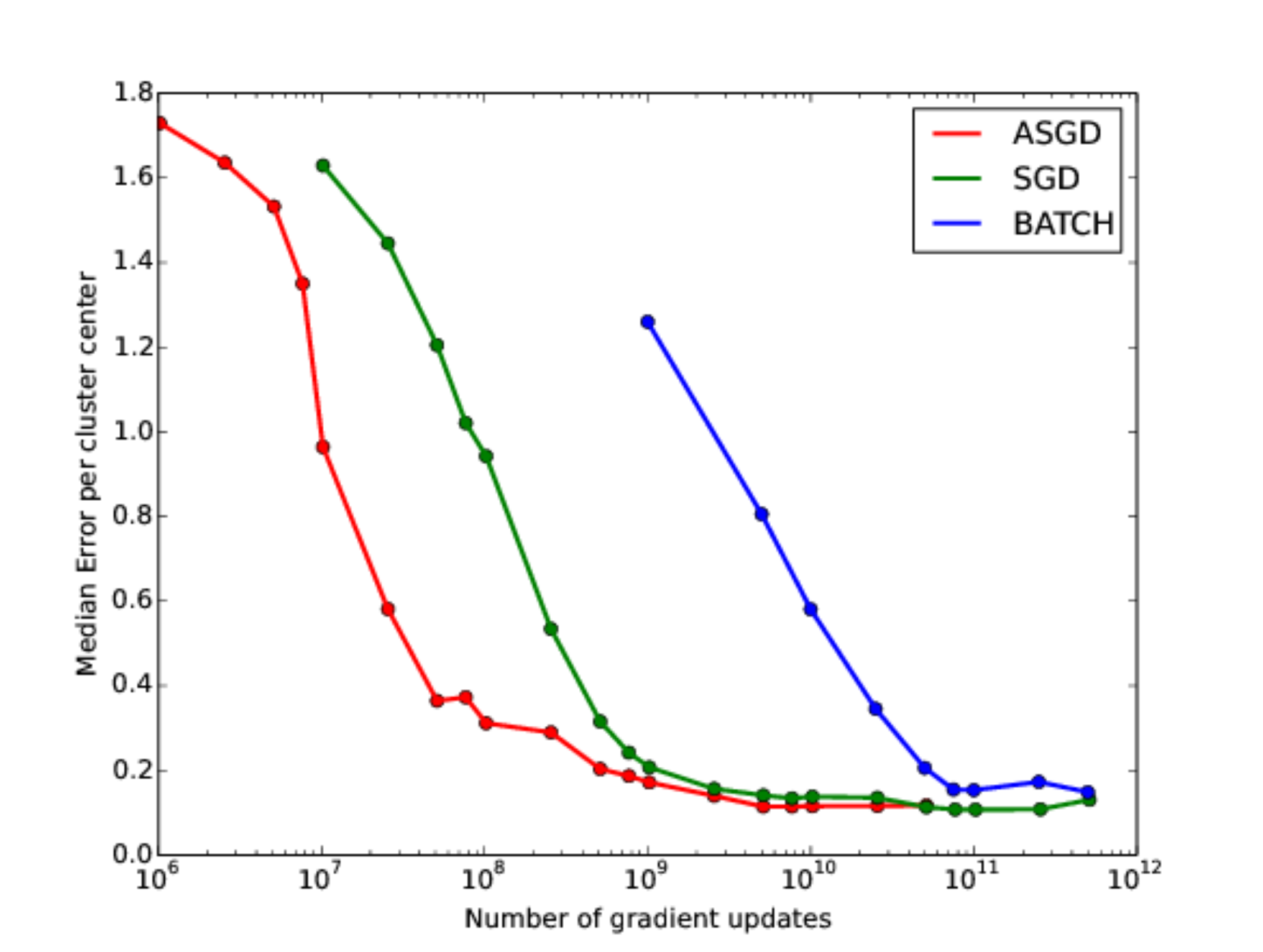}
\includegraphics[width=0.85\linewidth]{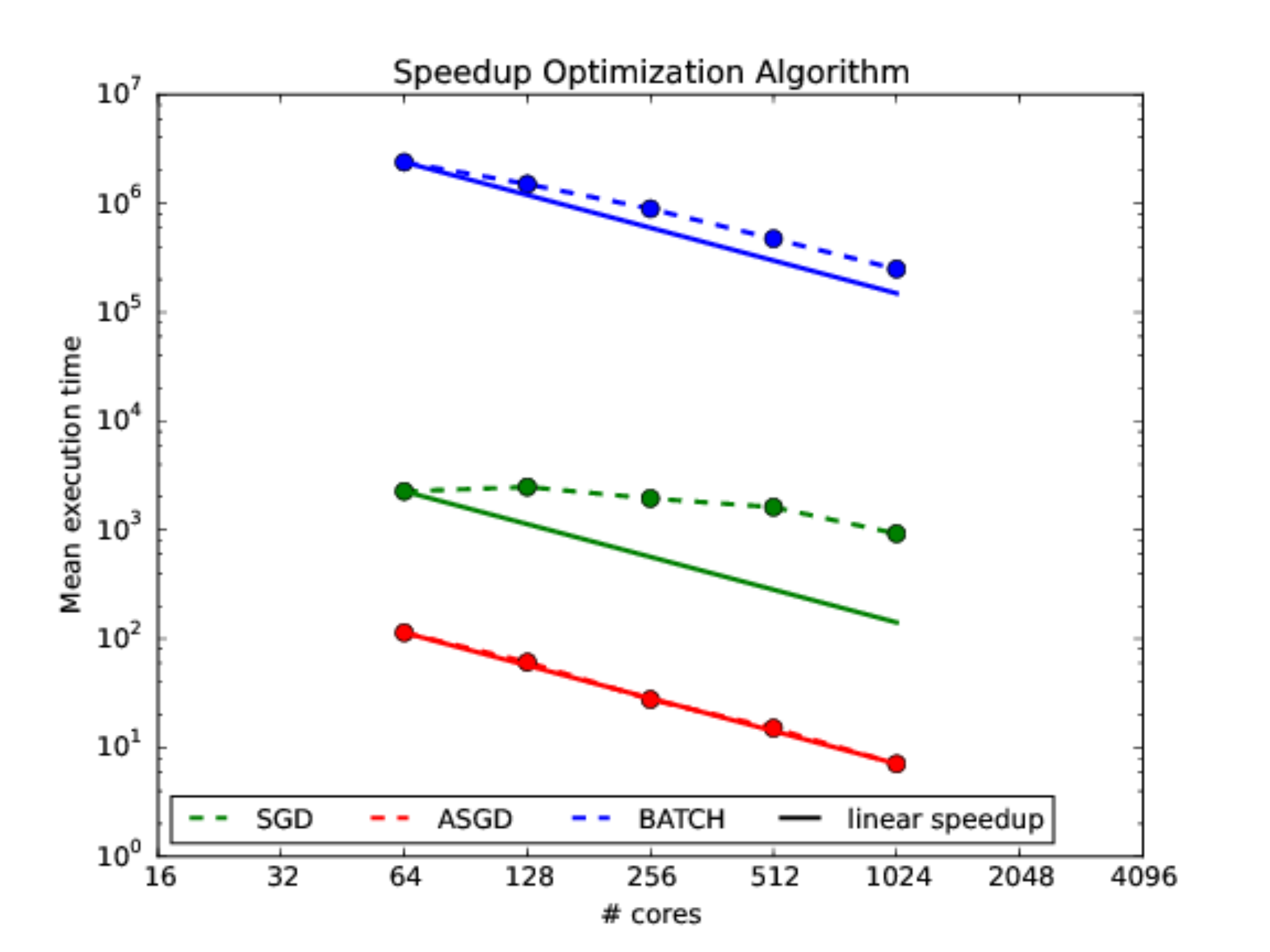}
\caption{\textbf{LEFT:}Convergence speed of different gradient descent methods used to solve
K-Means clustering
with $K=100$ on a $10$-dimensional target space parallelized  over $1024$ cores
on a cluster. The ASGD method outperforms communication free SGD
\cite{SGDsmola} and
MapReduce
based BATCH \cite{chu2007map} optimization by the order of a magnitude.
\textbf{RIGHT:} Scaling properties of the same experiment: ASGD is not only faster 
than SGD and BATCH, it also provides linear strong scaling in the number of CPUs, while
SGD suffers from increasing communication overheads. See \cite{arxiv} for detailed 
evaluations. 
\label{fig_eval_conv}
}
\end{figure}

In \cite{arxiv}, we introduced a lock-free parallelization
method for the computation of stochastic gradient optimization of large scale
machine learning algorithms, which is based on the asynchronous communication
paradigm. Figure \ref{fig_eval_conv} displays the key results of \cite{arxiv},
showing that Asynchronous Stochastic Gradient Descent (ASGD) outperforms
both batch and online algorithms in terms of convergence speed, scalability 
and prediction error rates.

In this paper, we extend the ASGD by an algorithm 
which automatically sets the communication and update frequencies. These are 
key parameters which have a large impact on the convergence performance. In 
\cite{arxiv}, they were set experimentally. However, they optimal choice is subject to
a large number of factors and influenced by the computing environment:
interconnection bandwidth, number nodes, cores per node, or NUMA layout, just 
to name a few. Hence, for a heterogeneous setup (e.g. in the Cloud) it is hardly possible to 
determine a globally optimal set of parameters. We therefore introduce a 
adaptive algorithm, which choses the parameters dynamically during the runtime of ASGD.         

\subsection{Related Work} 
Recently, several approaches towards an effective parallelization
of the SGD optimization have been proposed. A detailed overview and in-depth
analysis of their application to machine learning can be found in \cite{SGDsmola}.\\
In this section, we focus on a brief discussion of related publications,
which provided the essentials for the ASGD approach:
\begin{itemize}
\item A theoretical framework for the analysis
of SGD parallelization performance has been presented in \cite{SGDsmola}.
The same paper also introduced a novel approach
(called SimuParallelSGD), which
avoids communication and any locking mechanisms up to a single and final
MapReduce step.
\item  A widely noticed approach for a
``lock-free'' parallelization of SGD on shared memory systems has been introduced 
in \cite{recht2011hogwild}. The basic idea
of this method is to explicitly ignore potential data races and to write
updates directly into the memory of other processes. Given a minimum level
of sparsity, they were able to show that possible data races will neither harm
the convergence nor the accuracy of a parallel SGD. Even more, without
any locking overhead, \cite{recht2011hogwild} sets the current performance
standard for shared memory systems.
\item In \cite{grunewald2013gaspi}, the concept of a  Partitioned
Global Address Space programming framework (called GASPI) has been introduced.
This provides an asynchronous,
single-sided communication
and parallelization scheme for cluster environments.
We build our asynchronous communication on the basis of this framework.
\end{itemize}

\section{The Machine Learning Optimization Problem}
From a strongly simplified perspective, machine learning tasks are usually solving 
the problem of  
inferring generalized models from a given dataset $X=\{x_0,\dots,x_m\}$ with $x_i
\in\mathbb{R}^n$,
which in case of supervised learning is also assigned with semantic labels
$Y=\{y_0,\dots,y_m\}, y_i\in\mathbb{R}$.
During the learning process, the quality of a model is
evaluated by use of so-called loss-functions, which measure how well the current model 
represents the given data. We write $x_j(w)$ or $(x_j,y_j)(w)$ to indicate the
loss of a data point for the current parameter set $w$ of the model function.
We will also refer to $w$ as the ``state'' of the model. The actual learning
is then the process of minimizing the loss over all samples.
This is usually done by a gradient descent over the partial derivative of
the loss function in the parameter space of $w$.

\subsubsection*{Stochastic Gradient Descent.}
Although some properties of Stochastic Gradient Descent approaches might  
prevent their successful application to some optimization
domains, they are well established in the machine learning community
\cite{bottou2010large}.
\begin{algorithm}
\caption{SGD with samples $X=\{x_0,\dots,x_m\}$, iterations $T$, steps size
$\epsilon$ and states $w$}
\label{algo_SGD}
\begin{algorithmic}[1]
\Require{$\epsilon>0$}
\ForAll{$t=0\dots T$ }
\State{\begin{bf}draw\end{bf} $j \in \{1\dots m\}$ uniformly at random}
\State{\begin{bf}update\end{bf} $w_{t+1} \leftarrow w_{t} - \epsilon\partial_wx_j(w_{t})$}
\EndFor
\State{\Return $w_T$}
\end{algorithmic}
\end{algorithm}
Following the notation in \cite{SGDsmola}, SGD can be formalized
in pseudo code as outlined in algorithm \ref{algo_SGD}.
For further simplification, we will write $\Delta_j(w_t) := \partial_wx_j(w_{t})$
for the remainder of this paper.

\subsection{Asynchronous SGD}\label{sec_ASGD_concept}
The basic idea of the ASGD algorithm is to port the ``lock-free'' shared memory approach from 
\cite{recht2011hogwild} to distributed memory systems. This is far from trivial,
mostly because communication latencies in such systems will inevitably cause 
expensive dependency locks if the communication is performed in common two-sided
protocols (such as MPI message passing or MapReduce). This is also the
motivation for SimuParallelSGD \cite{SGDsmola} to avoid communication
 during the optimization: locking costs are usually much higher than the information
gain induced by the communication.\\
We overcome this dilemma by the application of the asynchronous, single-sided 
communication model provided by \cite{grunewald2013gaspi}: individual processes
send mini-BATCH \cite{sculley2010web} updates completely uninformed 
of the recipients status whenever they are ready to do so. On the recipient 
side, available updates are included in the local computation as available. 
In this scheme, no process ever waits for any communication to be sent or 
received. Hence, communication is literally ``free'' (in terms of latency).\\
Of course, such a communication scheme will cause data races and race conditions: 
updates might  be (partially) overwritten before they were used or even might be 
contra productive because the sender state is way behind the state of the recipient.\\
ASGD solves these problems by two strategies: first, we obey 
the sparsity requirements introduced by \cite{recht2011hogwild}.
This can be 
achieved by sending only partial updates to a few random recipients. Second,
we introduced a Parzen-window function, selecting only those updates
for local descent which are likely to improve the local state.          
ASGD is formalized and implemented on the basis of the SGD 
parallelization presented in \cite{SGDsmola}. In fact, the asynchronous communication is 
just added to the existing approach. This is based on the assumption that communication 
(if performed correctly) can only improve the gradient descent - especially when it 
is ``free''. If the communication interval is set to infinity, ASGD will become 
SimuParallelSGD.   
\subsubsection*{Implementation.}
Our ASGD implementation is based on the open-source library 
GPI 2.0\footnote{Download available at http://www.gpi-site.com/gpi2/},
which provides a C++ interface to the GASPI specification.
\subsubsection*{Parameters.}
ASGD takes several parameters, which can have a strong influence on the convergence 
speed and quality:
$\bf T$ defines the size of the data partition for each thread,
$\bf \epsilon$ sets the gradient step size (which needs to be fixed following
the theoretic constraints shown in \cite{SGDsmola}),
$\bf b$ sets the size of the mini-batch aggregation, and
$\bf I$ gives the number of SGD iterations for each thread. Practically, this also equals the number 
of data points touched by each thread.
\subsubsection*{Initialization.} 
The initialization step is straight forward and analog to SimuParallelSGD \cite{SGDsmola}
: the data is split into working packages of size $T$ and distributed to the 
worker threads. A control thread generates initial, problem dependent values for $w_0$
and communicates $w_0$ to all workers. From that point on, all workers run 
independently.
It should be noted, that $w_0$ also could be initialized with the 
preliminary results of a previously early terminated optimization run.      
\subsubsection*{Updating.}
The online gradient descent update step is the key leverage point of the ASGD 
algorithm. The local state $w^i_t$ of thread $i$ at iteration $t$ is updated 
by an externally modified step $\overline{\Delta_t(w^i_{t+1})}$, which not
only depends on the local $\Delta_t(w^i_{t+1})$ but also on a possible
communicated state $w^j_{t'}$ from an unknown iteration $t'$ at some random thread $j$:   
\begin{equation}
\overline{\Delta_t(w^i_{t+1})}=w^i_t-{1\over 2}\left( w^i_t + w^j_{t'} \right) + \Delta_t(w^i_{t+1})
\label{eq_ASGD_1}
\end{equation}
Figure \ref{fig_ASGD_pWindow} gives a schematic overview 
of the update process.
\subsubsection*{{Parzen-Window} Optimization.}
\begin{figure}[t]
\includegraphics[width=\textwidth]{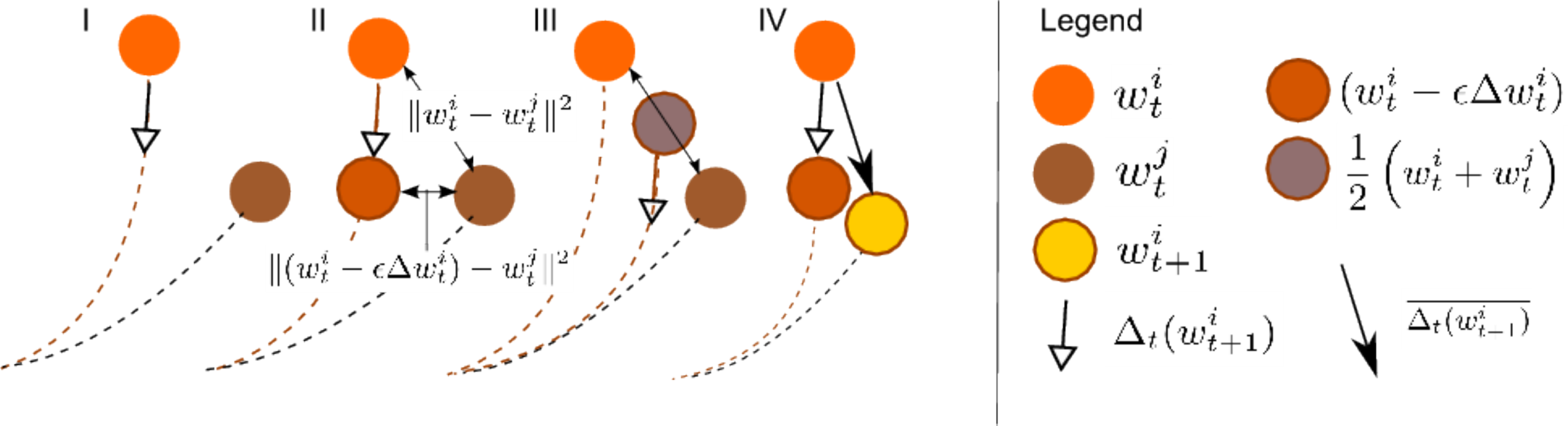}
\caption{ASGD updating. {\normalfont This figure visualizes the update algorithm 
of a process with state $w^i_t$, its local mini-batch update 
$\Delta_t(w^i_{t+1})$ and received external state  $w^j_t$ for a 
simplified 1-dimensional optimization problem. The dotted lines indicate 
a projection of the expected descent path to an (local) optimum. 
{\bf I:} Initial setting: ${\bf \Delta}_M(w^i_{t+1})$ is computed and $w^j_t$ is in the external
buffer.
{\bf II:} Parzen-window masking of $w^j_t$. Only if the condition of equation 
(\ref{eq_parzen}) is
met, $w^j_t$ will contribute to the local update.
{\bf III:} Computing $\overline{{\bf\Delta}_M(w^i_{t+1})}$.
{\bf IV:} Updating $w^i_{t+1} \leftarrow w^i_t -\epsilon\overline{{\bf\Delta}_M
(w^i_{t+1})}$.   
}
\label{fig_ASGD_pWindow}
}
\end{figure}

As discussed before, 
the asynchronous communication scheme is prone 
to cause data races and other conditions during the update. Hence, we introduced a 
Parzen-window like function $\delta(i,j)$ to avoid ``bad'' update conditions. 
The handling of data races is discussed in \cite{arxiv}. 
\begin{equation}
\delta(i,j) := \left\{
\begin{tabular}{ll}
  $1$ & if $\|(w^i_t-\epsilon\Delta w^i_t)-w^j_{t'}\|^2 < \|w^i_t-w^j_{t'}\|^2$ \\
  $0$ & otherwise 
\end{tabular}
\right.,
\label{eq_parzen}
\end{equation}
We consider an update to be ``bad'', if the external state $w^j_{t'}$ would direct
the update away from the projected solution, rather than towards it. Figure
\ref{fig_ASGD_pWindow} shows the evaluation of $\delta(i,j)$,
which is then plugged into the update functions of ASGD in order to exclude 
undesirable external states from the computation. Hence, equation (\ref{eq_ASGD_1}) 
turns into
\begin{equation}
\overline{\Delta_t(w^i_{t+1})}=\left[w^i_t-{1\over 2}\left( w^i_t + w^j_{t'} \right)\right]
\delta(i,j) + \Delta_t(w^i_{t+1})
\label{eq_ASGD_1_parzen}
\end{equation}

In addition to the Parzen-window, we also introduced a mini-batch update
in \cite{arxiv}: instead of updating after each step, several updates are aggregated 
into mini-batches of size $b$.
We are writing ${\bf\Delta}_M$ in order to differentiate mini-batch steps
from single sample steps $\Delta_t$ of sample $x_t$:   
\begin{equation}
\overline{{\bf\Delta}_M(w^i_{t+1})}=\left[w^i_t-{1\over 2}\left( w^i_t + w^j_t \right)\right]
\delta(i,j) + 
{\bf\Delta}_M(w^i_{t+1})
\label{eq_ASGD_miniB}
\end{equation}

\subsubsection*{Computational Costs of Communication.}
Obviously, the evaluation of $\delta(i,j)$ comes at some computational cost.
Since $\delta(i,j)$
has to be evaluated for each received message,
the ``free'' communication is actually not so free after all.
However, the costs are very low and can be reduced  
 to the computation of the distance between two states, which can be
achieved linearly in the dimensionality of the parameter-space of $w$ and the mini-batch size: 
$O({1\over b}|w|)$. 
In practice, the communication frequency ${1\over b}$ is mostly constrained by the 
network bandwidth and latency between the compute nodes, which 
is subject to our proposed automatic adaption algorithm in the next section.
\ref{sec_balance}.    

\subsubsection*{The ASGD Algorithm.}
Following \cite{arxiv}, the final ASGD algorithm 
with mini-batch size $b$, number of iterations $I$ 
and learning rate $\epsilon$ can be implemented as shown in
algorithm \ref{algo_ASGD}. 
\begin{algorithm}
\caption{ASGD $(X=\{x_0,\dots,x_m\},T,\epsilon,w_0,b)$}
\label{algo_ASGD}
\begin{algorithmic}[1]
\Require{$\epsilon>0, n>1$}
\State{\begin{bf}define\end{bf} $H=\lfloor {m\over n}\rfloor$}
\State{randomly \begin{bf}partition\end{bf} $X$, giving $H$ samples to each node}
\ForAll{$i \in \{1,\dots,n\}$ \begin{bf}parallel\end{bf} }
\State{randomly {\bf shuffle} samples on node $i$}
\State{\begin{bf}init\end{bf} $w^i_{0}=0$}
\ForAll{$t=0\dots T$ }
\State{\begin{bf}draw\end{bf} mini-batch $M \leftarrow b$ samples from $X$}
\State{\begin{bf}update\end{bf} $w^i_{t+1} \leftarrow w^i_{t} - \epsilon\overline{{\bf\Delta}_M(w^i_{t+1})}$}
\State{\begin{bf}send\end{bf} $w^i_{t+1}$ to random node $\neq i$}
\EndFor
\EndFor
\State{\Return  $w^1_I$}
\end{algorithmic}

\end{algorithm}
At termination, all nodes $w^i, i \in \{1,\dots,n\}$ hold small local
variations of the global
result. We simply return one of these (namely $w^1_I$). Experiments showed,
that further aggregation of the $w^i_I$ (via map reduce) provides no
improvement of the results and can be
neglected.

\section{Communication load balancing}\label{sec_balance}
\begin{figure}[t]
\centering
\includegraphics[width=0.43\textwidth]{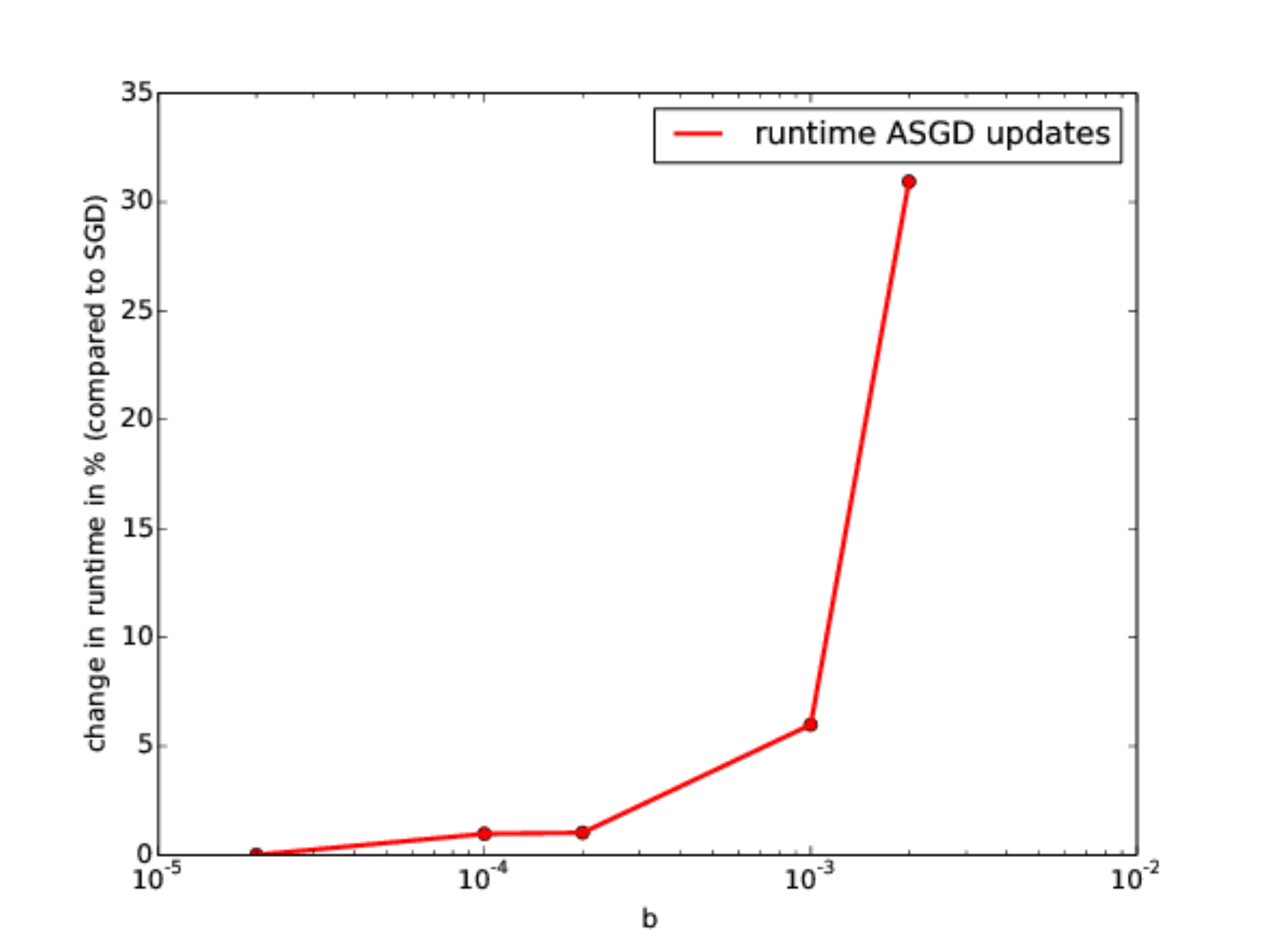}
\includegraphics[width=0.43\textwidth]{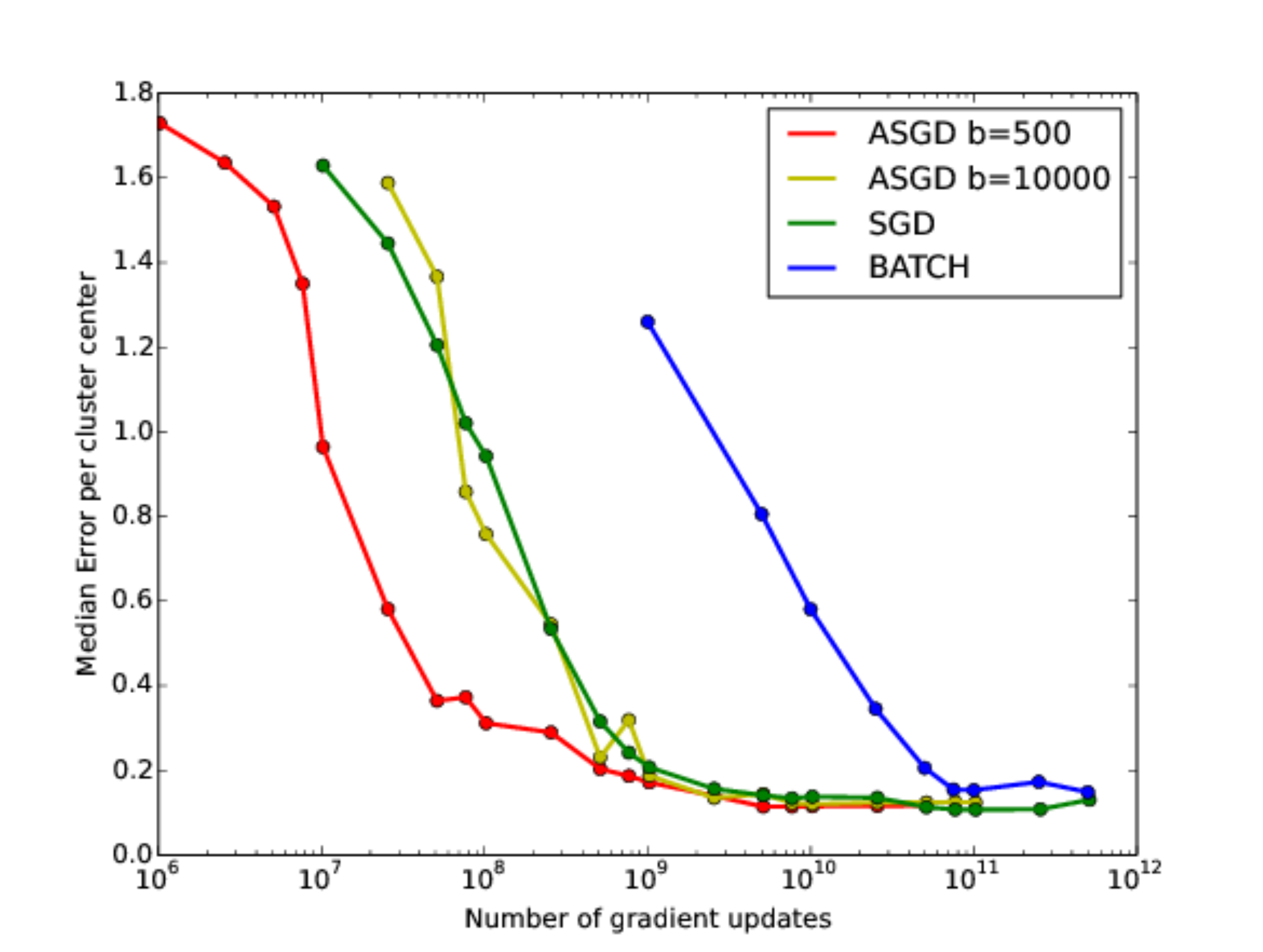}
\caption{{\bf LEFT:} Communication cost of ASGD. The cost of higher communication 
frequencies $1\over b$ in ASGD updates compared to communication free SGD updates. 
{\bf RIGHT:} Convergence speed of ASGD with a communication frequencies of $1\over 100000$ 
compared to $1\over 500$ in relation to the other methods. Results on Synthetic 
data with $D=10, K=100$.
\label{fig_eval_b}}
\end{figure}
The impact of the communication frequencies of $1\over b$ on the
convergence properties of ASGD are displayed in figure \ref{fig_eval_b}. 
If the frequency is set to lower values, the convergence 
moves towards the original SimuParallelSGD behavior.
The results in Figure \ref{fig_eval_b} show that the choice of the communication frequency 
${1\over b}$ has a significant impact on the convergence speed.
Theoretically, more communication should be beneficial. However, due to
the limited bandwidth, the practical limit is expected to be far from $b=1$.\\ 
The choice of an optimal $b$ strongly depends on the data (in terms of dimensionality) 
and the computing environment:
interconnection bandwidth and latency, number of nodes, cores per node, NUMA layout and 
so on.\\
In \cite{arxiv}, the ASGD approach has only been tested in an HPC cluster 
environment with Infiniband interconnections, where neither bandwidth nor latency
issues were found to have significant effect on the experiments. Hence, $b$ was set
to a fixed value which has been selected experimentally.\\
However, for most Big Data applications, especially in HTC environments like the cloud, 
Infiniband networks are not very common. Instead, one usually has to get
along with Gigabit-Ethernet connections, which even might suffer from external 
traffic. Figures \ref{fig_eval_large} and \ref{fig_eval_large_message} show
the effect of reduced bandwidth and higher latencies on the ASGD performance:
As one can expect, the number of messages that can be passed through the network,  
underlies stronger bounds compared to Infiniband. Notably, our experiments indicate, 
that there appears to be a clear local optimum for $b$, where the number of messages 
correlates to the available bandwidth. Because this local optimum might even change during
runtime (through external network traffic) and a full series of experiments on 
large datasets (in order to determine $b$) is anything but practical, we
propose an adaptive algorithm which regulates $b$ during the runtime of ASGD.\\

\subsection{Adaptive optimal $b$ estimation}
The GPI2.0 interface allows the monitoring of outgoing asynchronous communication 
queues. By keeping a small statistic over the past iterations, our approach
dynamically increases the frequency $1\over b$ when queues are running low, and
decreases or holds otherwise. Algorithm \ref{algo_aB} is run on all nodes independently,
dynamically setting $b$ for all local threads.\\
            
\begin{algorithm}
\caption{adaptiveB $(q_{opt},q_0,q_1,q_2,\gamma)$}
\label{algo_aB}
\begin{algorithmic}[1]
\State{\begin{bf}get current queue state\end{bf} $q_0$}
\State{\begin{bf}compute gradient\end{bf} $\Delta_q=(q_{opt}-q_0)-(q_2-q_0)$}
\State{\begin{bf}update \end{bf} $ b=b-\Delta_q*\gamma$}
\State{\begin{bf}update history\end{bf} $ q_2=q1, q_1=q_0$}
\State{\Return  $b$}
\end{algorithmic}
\end{algorithm}
Where $q_0$ is the current queue size queried from the GPI interface, $q_{opt}$
the target queue size, $q_1,q_2$ the queue history and $\gamma$ the step size
regularisation.  

\section{Experiments} \label{sec_eval}
We evaluate the performance of our proposed method in terms of convergence speed, 
scalability and error rates of the learning objective function using the 
K-Means Clustering algorithm. The motivation to choose this algorithm for
evaluation is twofold: First, K-Means is probably one of the simplest machine
learning algorithms known in the literature (refer to \cite{jain2010data} for a 
comprehensive overview). This leaves little room for algorithmic optimization 
other than the choice of the numerical optimization method. Second, it is also
one of the most popular\footnote{The original paper \cite{lloyd1982least} has been
cited several thousand times.} unsupervised learning algorithms with a wide 
range of applications and a large practical impact.        

\subsection{K-Means Clustering}
K-Means is an unsupervised learning algorithm, which tries to find the
underlying cluster structure of an unlabeled vectorized dataset.   
Given a set of $m$ $n$-dimensional points $X=\{x_i\},i=1,\dots,m$, which is to
be clustered into a set of $k$ clusters, $w=\{w_k\},k=1,\dots,k$. The K-Means
algorithm finds a partition such that the squared error between the
empirical mean of a cluster and the points in the cluster is minimized.\\
%Let $\mu_k$ be the mean of cluster $c_k$. The squared error between
%$\mu_k$ and the points in cluster $c_k$ is then defined as \cite{jain2010data}:
%\begin{equation}
%J(c_k) = \sum_{x_i\in c_k}\|x_i-\mu_k\|^2
%\end{equation}
%The objective of the K-Means algorithm is then to minimize the squared error
%over all clusters:
%\begin{equation}
%J(C) = \sum_{k=1}^K\sum_{x_i\in c_k}\|x_i-\mu_k\|^2
%\end{equation} 
It should be noted, that finding the global minimum of the squared error
over all $k$ clusters $E(w)$ is proven to be 
a NP-HARD problem \cite{jain2010data}. Hence, all optimization methods 
investigated in this paper are only approximations of an optimal solution.   
However, it has been shown \cite{Meila}, that K-Means finds local optima 
which are very likely to be in close proximity to the global minimum if the 
assumed structure of $k$ clusters is actually present in the given data.  

\subsubsection*{Gradient Descent Optimization}
Following the notation given in \cite{bottou1994convergence}, K-Means is 
formalized as minimization problem of the quantization error $E(w)$: 
\begin{equation}
E(w)=\sum_i{1\over 2}(x_i-w_{s_i(w)})^2,
\label{eq_kmeans}
\end{equation} 
where $w=\{w_k\}$ is the target set of $k$ prototypes for given $m$ examples 
$\{x_i\}$ and $s_i(w)$ returns the index of the closest prototype to the sample 
$x_i$.
The gradient descent of the quantization error $E(w)$ is then derived as 
$\Delta(w)={\partial E(w)\over \partial w}$. For the usage with the 
previously defined gradient descent algorithms, 
this can be reformulated to the following update
function with step size $\epsilon$. 
\begin{equation}
\Delta(w_k)=\left\{
\begin{tabular}{ll}
  $x_i-w_k$ & if $k=s_i(w)$ \\
  $0$ & otherwise 
\end{tabular}
\right.
\label{eq_km_online}
\end{equation}

\subsection{Setup}
The experiments were conducted on a Linux cluster with a BeeGFS\footnote{see www.beegfs.com for details}
parallel file system.  
Each compute node is equipped
 with dual Intel Xeon E5-2670, totaling to 16 cores per node, 32 GB RAM and 
interconnected with FDR Infiniband or Gigabit-Ethernet.
If not noted otherwise, we used a standard of 64 nodes to compute the 
experimental results (which totals to 1024 CPUs). 

\subsubsection*{Synthetic Data Sets}
The need to use synthetic datasets for evaluation arises from several rather 
profound reasons: (I) the optimal solution is usually unknown for real data,
(II) only a few very large datasets are publicly available, and, (III) we even 
need a collection of datasets with varying parameters such as dimensionality $n$,
size $m$ and number of clusters $k$ in order to evaluate the scalability.\\
The generation of the data follows a simple heuristic: given $n,m$ and $k$ we
randomly sample $k$ cluster centers and then randomly draw $m$ samples. Each
sample is randomly drawn from a distribution which is uniquely generated for 
the individual centers. Possible cluster overlaps are controlled by additional
minimum cluster distance and cluster variance parameters. The detailed properties 
of the datasets are given in the context of the experiments.     
\subsubsection*{Evaluation}
Due to the non-deterministic nature of stochastic methods and the fact that 
the investigated K-Means algorithms might get stuck in local minima, we  
apply a 10-fold evaluation of all experiments. If not noted otherwise, plots
show the median results. Since the variance is usually magnitudes lower than
the plotted scale, we neglect the display of variance bars in the plots for the 
sake of readability.\\ 
Errors reported for the synthetic datasets are computed as follows: We use the
``ground-truth'' cluster centers from the data generation step to measure their
distance to the centers returned by the investigated algorithms. 

\begin{figure}[ht]
\centering
\includegraphics[width=0.43\textwidth]{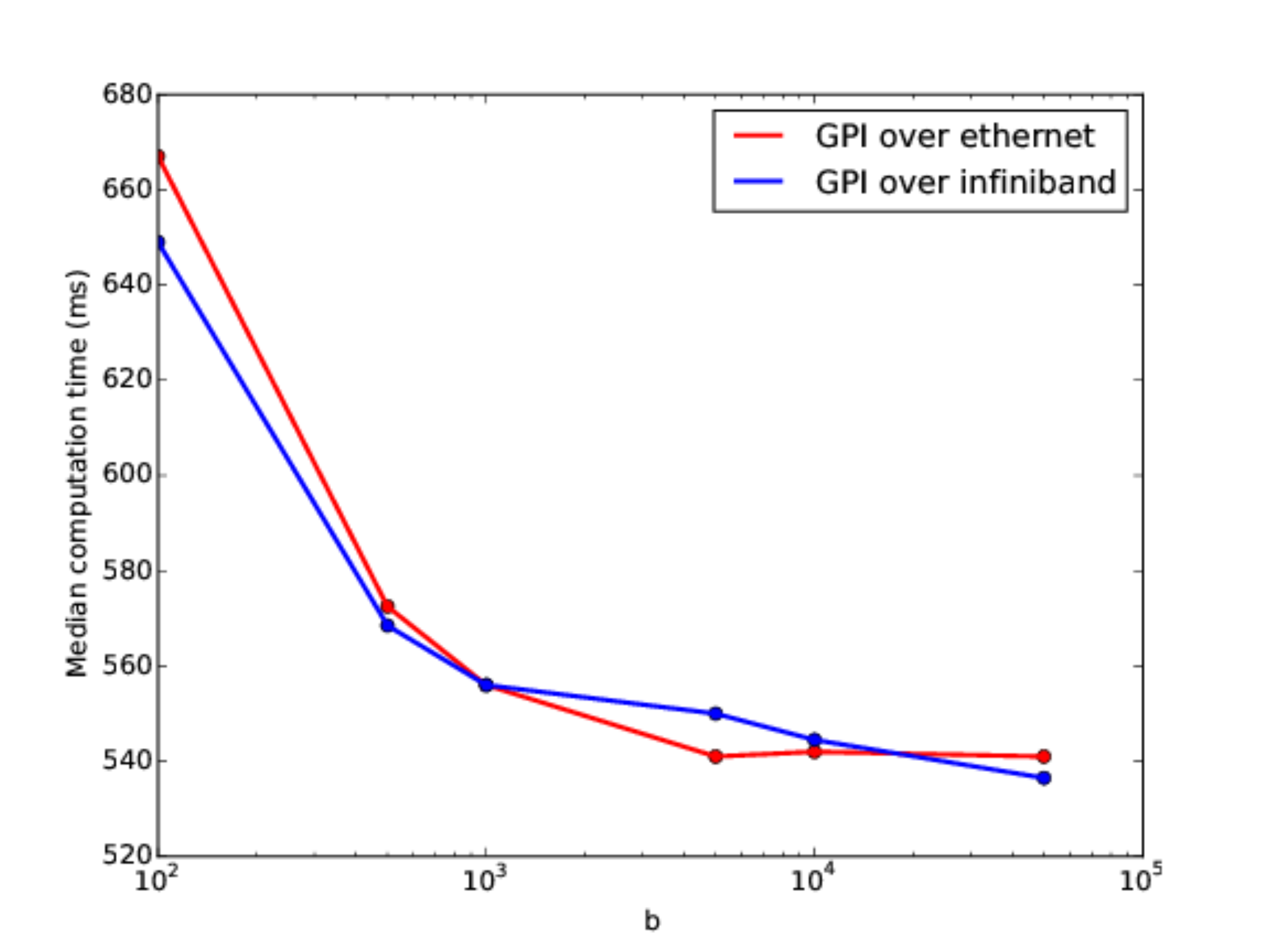}
\includegraphics[width=0.43\textwidth]{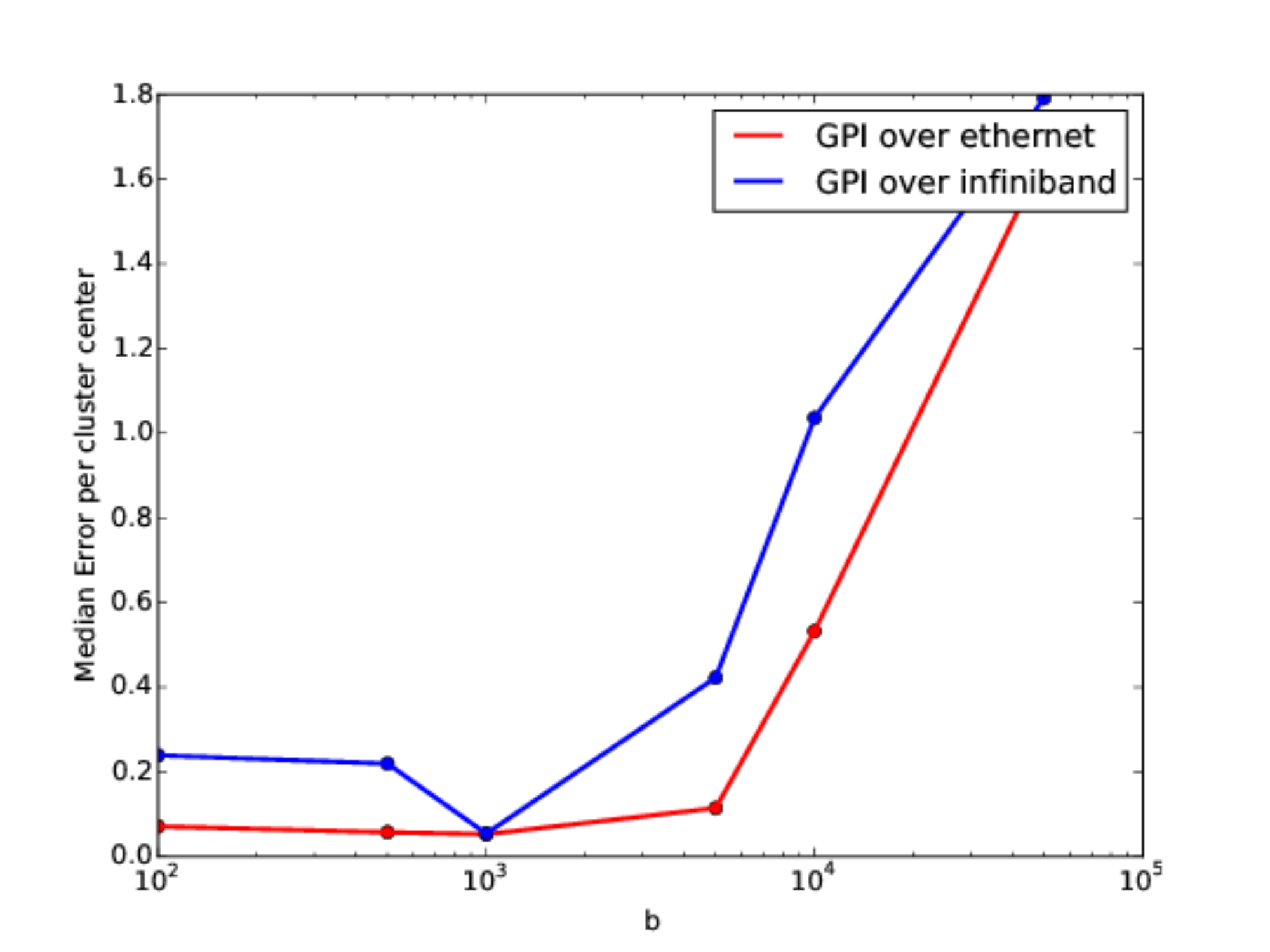}
\caption{Comparing ASGD performance on Gigabit-Ethernet vs. Infiniband 
interconnections.
{\bf LEFT:} Median runtime of ASGD for altering communication 
frequencies $1\over b$. 
{\bf RIGHT:} Median Error rates of ASGD
Results on Synthetic data with $D=10, K=10$, which results in small messages ($50$ byte) 
shows hardly any difference between the performance of both connection types.
\label{fig_eval_small}}
\end{figure}

\begin{figure}[ht]
\centering
\includegraphics[width=0.43\textwidth]{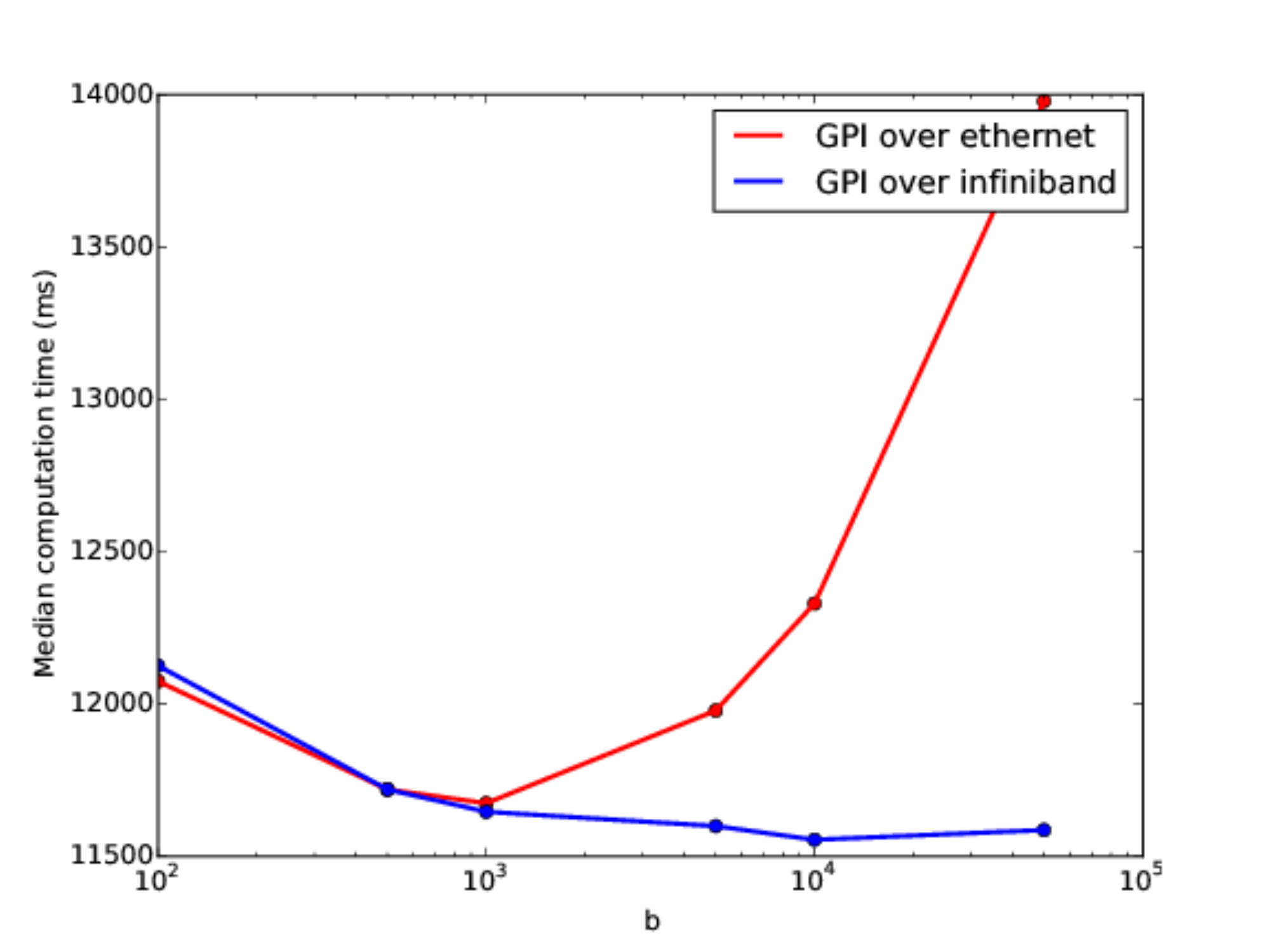}
\includegraphics[width=0.43\textwidth]{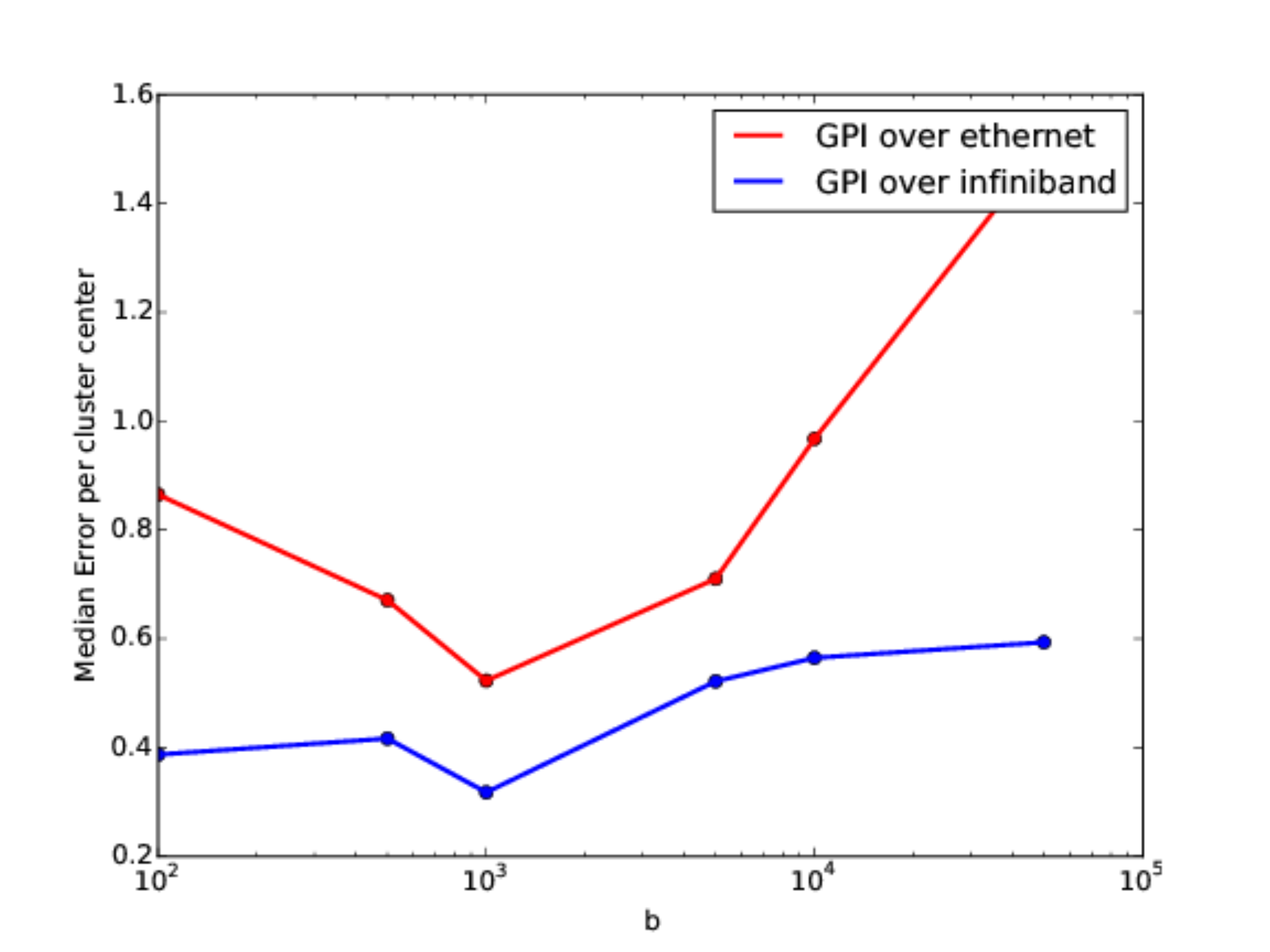}
\caption{Same experiments as in figure \ref{fig_eval_small}, but with a lager problem
($D=100, K=100$: message size $5$kB) shows significant differences in performance. Notably, the
Gigabit-Ethernet interconnect shows a local optimum for $b=1000$. 
\label{fig_eval_large}} 
\end{figure}

\begin{figure}[ht]
\centering
\includegraphics[width=0.43\textwidth]{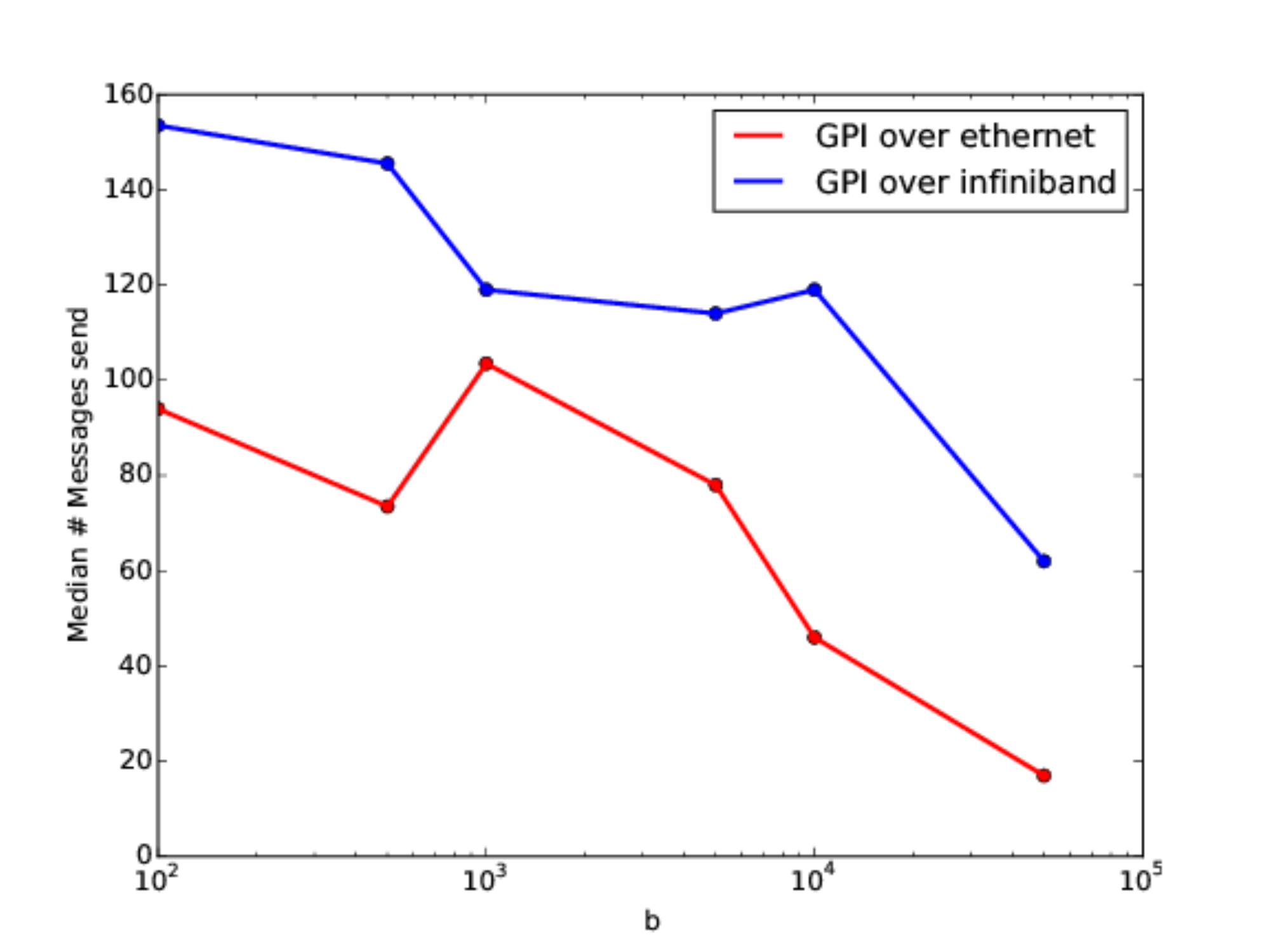}
\includegraphics[width=0.43\textwidth]{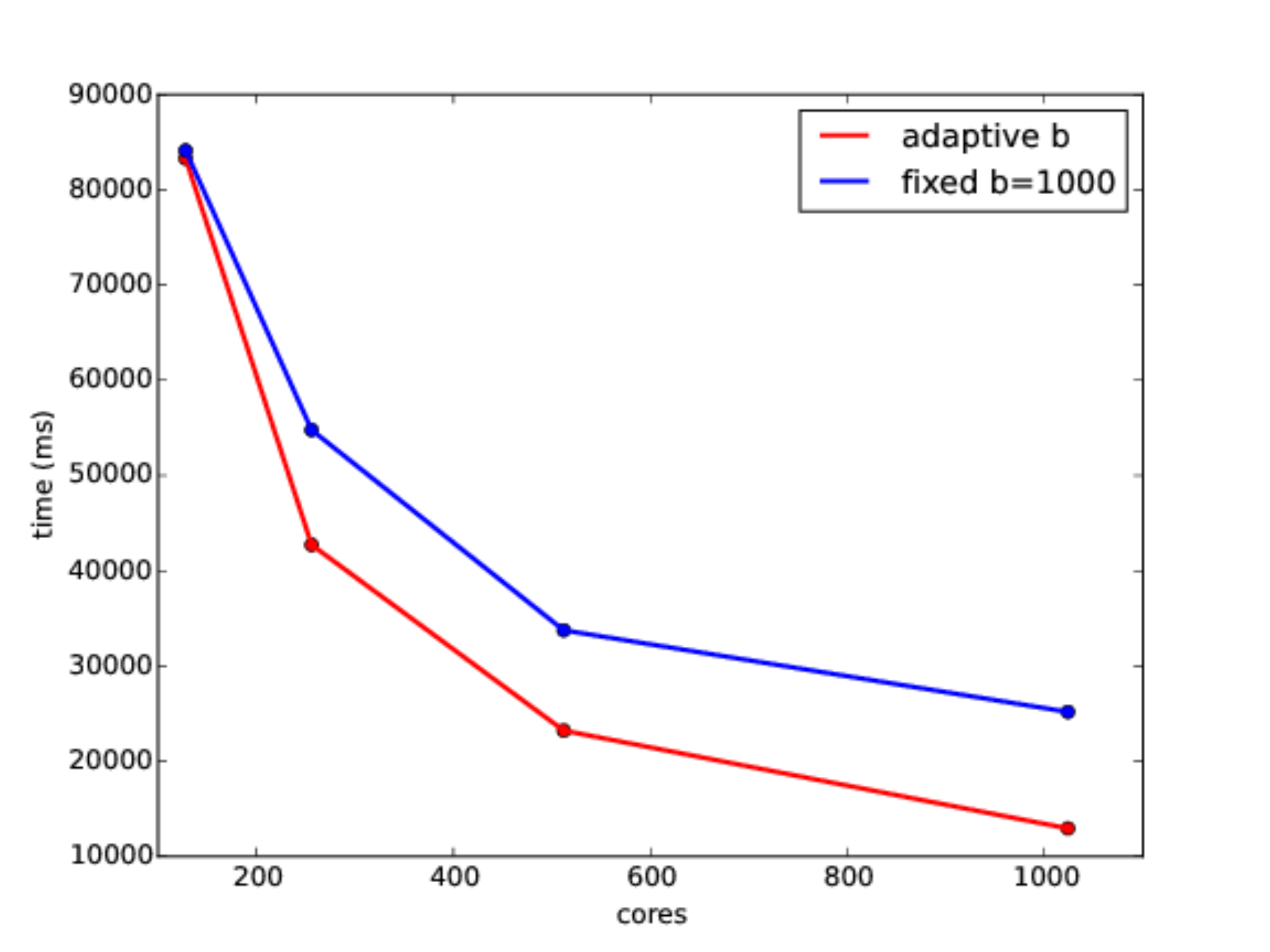}
\caption{{\bf LEFT:} Same experiments as in figure \ref{fig_eval_large},
showing the median number of ``good'' messages send. Again, the
Gigabit-Ethernet interconnect shows a local optimum for $b=1000$.
{\bf RIGHT:} Evaluation of the scaling properties of ASGD on Gigabit-Ethernet. 
Comparing a fixed $b$ and our new adaptive $b$ algorithm. 
\label{fig_eval_large_message}}
\end{figure}
\subsubsection*{Results.}
Figure \ref{fig_eval_small} shows that the performance of the ASGD algorithm for 
problems with small message sizes is hardly influenced by the network bandwidth.
Gigabit-Ethernet and Infiniband implementations have approximately the same performance.
This situation changes, when the message size in increased. Figure \ref{fig_eval_large}
shows that the performance is breaking down, as soon as the Gigabit-Ethernet connections
reach their bandwidth limit.\\
Figure \ref{fig_eval_large_message} shows that this effect can be softened by 
the usage of our adaptive message frequency algorithm, automatically selecting 
the current maximum frequency which will not exceed the available bandwidth. 
\section{Conclusions}
The introduced load balancing algorithm simplifies the usage of the ASGD optimization
algorithm in machine learning applications on HTC environments. 
\bibliographystyle{abbrv}
\bibliography{ASGD}  % sigproc.bib is the name of the Bibliography in this case
\end{document}